\documentclass[twoside]{LCWS11}
\usepackage[latin1]{inputenc}
\usepackage[dvips]{graphicx,epsfig,color}
\usepackage{wrapfig,rotating}
\usepackage{amssymb,amsmath,array}

\newcommand{\gsim}{\mbox{ \raisebox{-1.0ex}{$\stackrel{\textstyle >}
{\textstyle \sim}$ }}}
\newcommand{\lsim}{\mbox{ \raisebox{-1.0ex}{$\stackrel{\textstyle <}
{\textstyle \sim}$ }}}
\newcommand{\aneq}{\!\!\!&=&\!\!\!} 
  \def\gtsim{\mathrel{\hbox{\raise0.2ex
\hbox{$>$}\kern-0.75em\raise-0.9ex\hbox{$\sim$}}}}
\def\ltsim{\mathrel{\hbox{\raise0.2ex
\hbox{$<$}\kern-0.75em\raise-0.9ex\hbox{$\sim$}}}}

\begin{document}
\title{
Decoupling Property of SUSY Extended Higgs Sectors and Implication for
Electroweak Baryogenesis} 
\author{Shinya Kanemura$^1$, Eibun Senaha$^2$ and Tetsuo Shindou$^3$
\vspace{.3cm}\\
1- Department of Physics, University of Toyama\\
 3190 Gofuku, Toyama 930-8555, Japan 
\vspace{.1cm}\\
2- Korea Institute for Advanced Study, School of Physics\\
 85 Hoegiro, Dongdaemun-gu, Seoul 130-722, Korea
\vspace{.1cm}\\
3- Division of Liberal Arts, Kogakuin University\\ 
    1-24-2 Shinjuku, Tokyo 163-8677, Japan\\
}

\maketitle

\begin{abstract}
 One-loop contributions to the Higgs potential at finite temperatures are
 discussed in the supersymmetric standard model with four Higgs doublet
 chiral superfields as well as a pair of charged singlet chiral superfields.
 The mass of the lightest Higgs boson $h$ is determined only by the
 D-term at the tree-level in this model, while the triple Higgs boson
 coupling for $hhh$ can receive a significant radiative correction. 
 The same nondecoupling effect can also contribute to realize the sufficient 
 first order electroweak phase transition, which is required for a
 successful scenario of electroweak baryogenesis. 
 This model can be a new candidate for a model in which the baryon
 asymmetry of the Universe is explained at the electroweak scale.
 We also discuss the implication for the measurement of the triple Higgs
 boson coupling at the ILC.
\end{abstract}

\section{Introduction}

The observed baryon asymmetry of the Universe is expressed by  
$n_b/n_\gamma \simeq (5.1-6.5) \times 10^{-10}$
 at the 95 \% CL~\cite{BAU}, where
$n_b$ is the difference in number density between baryons and
anti-baryons and $n_\gamma$ is the number density of photons.  
It is known that to generate the baryon asymmetry
from the baryon symmetric world, the Sakharov's conditions
have to be satisfied~\cite{sakharov}:
1) Baryon number nonconservation, 
2) C and CP violation, 
3) Departure from the thermal equilibrium.
The electroweak gauge theory can naturally satisfy the above three
conditions, where the third condition is satisfied when 
the electroweak phase transition (EWPT) is of strongly first order.
This scenario is often called the electroweak baryogenesis~\cite{B-EW}.
The scenario is necessarily related to the Higgs boson dynamics. 
It is directly testable at collider experiments.
In the standard model (SM), however, the CP violation by the
Cabibbo-Kobayashi-Maskawa matrix is quantitatively
insufficient~\cite{ewbg_sm_cp}. 
In addition, from the LEP direct search results
the requirement of sufficiently strong first order EWPT 
requires  a light Higgs boson whose mass is too small to satisfy 
the constraint~\cite{h-search-LEP}.
A viable model for successful electroweak baryogenesis would
be the two Higgs doublet model (THDM)~\cite{B-EW3}.
In Ref.~\cite{ewbg-thdm2}, the connection between the first order EWPT
and the triple coupling for the SM-like Higgs boson $h$ (the $hhh$ coupling)
has been clarified. In the model with sufficiently strong first order EWPT,
the $hhh$ coupling constant significantly deviates from the SM
prediction due to the same nondecoupling quantum effects
of additional scalar bosons.
Such nondecoupling effects on the $hhh$ coupling constant have been
studied in Ref.~\cite{KOSY}.

Supersymmetry is a good new physics candidate, which eliminates the
quadratic divergence in the one-loop calculation of the Higgs boson
mass. The lightest SUSY partner particle with the R
parity can naturally be a candidate for the cold dark matter.    
In the minimal supersymmetric SM (MSSM), there are many studies to realize
the electroweak baryogenesis~\cite{ewbg-mssm,Carena:2008vj,Funakubo:2009eg}. 
Currently, this scenario is highly constrained by the experimental
data. According to Ref. \cite{Carena:2008vj}, the strong first order
EWPT is possible if
$m_h\lsim 127$~GeV and $m_{\tilde{t}_1}\lsim 120$~GeV,
where $h$ is the lightest Higgs boson and $\tilde{t}_1$ is the lightest
stop.  
To satisfy the LEP bound on $m_h$, the soft SUSY breaking mass for the left-handed stop 
should be greater than 6.5~TeV. The most striking feature of this scenario is that
the electroweak vacuum is metastable and the global minimum is a
charge-color-breaking vacuum. 
Many studies on the electroweak baryogenesis
have been done in such singlet-extended MSSMs; i.e., 
the Next-to-MSSM~\cite{Funakubo:2005pu}, the nearly MSSM or the minimal
non-MSSM~\cite{EWPT_nMSSM}, the $U(1)'$-extended MSSM~\cite{EWPT_UMSSM},  
the secluded $U(1)'$-extended MSSM~\cite{Chiang:2009fs,Kang:2009rd},
and so on.
In the singlet-extended MSSM, however, the vacuum structure is inevitably more complicated
than the MSSM, giving rise to the unrealistic vacua 
in the large portion of the parameter space, especially electroweak baryogenesis-motivated
scenario~\cite{Funakubo:2005pu,Chiang:2009fs}.

In this talk, we discuss how the electroweak phase transition can be
of sufficiently strong first order in an extended SUSY standard model~\cite{kss}, where 
a pair of extra  doublet chiral superfields $H_3$ ($Y=-1/2$) and $H_4$ ($Y=+1/2$)
and a pair of charged singlet chiral superfields $\Omega_1$ ($Y=+1$) and
$\Omega_2$ ($Y=-1$) are introduced in addition to the MSSM content.
A motivation of this model is a SUSY extension of
the model to generate the tiny neutrino masses
by radiative corrections~\cite{SS-R,Aoki:2008av}.
In the present model,
there is no tree-level F-term contribution to the mass of the lightest Higgs boson,
but there can be large one-loop corrections to the triple Higgs boson
coupling due to the additional bosonic loop
contribution~\cite{ksy}.
Similarly to the case of the non-SUSY THDM, these nondecoupling
bosonic loop contributions can also make first order phase transition stronger.
We here show that the EWPT can be of sufficiently
strong first order in this model.

\section{Model}

In addition to the standard gauge symmetries,
we impose a discrete $Z_2$ symmetry for simplicity. 
Although the $Z_2$ symmetry is not essential for our discussion,
the symmetry works for avoiding the flavor changing neutral
current at the tree level~\cite{fcnc,barger,typeX,aksy}. 
Furthermore, we assume that there is the R parity in our model.
The superpotential is given by 
\begin{align}
 W &=(y_u)^{ij} U_i^c H_2\cdot Q_j+(y_d)^{ij} D_i^c H_1\cdot Q_j
 +(y_e)^{ij}E^c_iH_1\cdot L_j \nonumber \\
 &+ \lambda_1\Omega_1 H_1\cdot H_3 
 + \lambda_2\Omega_2 H_2\cdot H_4 
 -\mu H_1\cdot H_2 -\mu^\prime H_3\cdot H_4 - \mu_\Omega \Omega_1 \Omega_2.
 \end{align}
The soft-SUSY-breaking terms are given by 
\begin{align}
 &{\mathcal L}_{\rm soft} =
 -\frac{1}{2}(M_1 \tilde{B}\tilde{B}
             +M_2 \tilde{W}\tilde{W}
             +M_3 \tilde{G}\tilde{G})
 -\left\{
 (\tilde{M}_{\tilde{q}}^2)_{ij} \tilde{q}_{Li}^\dagger \tilde{q}_{Lj}  +
 (\tilde{M}_{\tilde{u}}^2)_{ij} \tilde{u}_{Ri}^\ast \tilde{u}_{Rj}
  \right. \nonumber\\
 &\left.
 +(\tilde{M}_{\tilde{d}}^2)_{ij} \tilde{d}_{Ri}^\ast \tilde{d}_{Rj}+
 (\tilde{M}_{\tilde{\ell}}^2)_{ij} \tilde{\ell}_{Li}^\dagger \tilde{\ell}_{Lj}  +
 (\tilde{M}_{\tilde{e}}^2)_{ij} \tilde{e}_{Ri}^\ast \tilde{e}_{Rj} 
 \right\} 
 -\left\{
   \tilde{M}_{H_1}^2\Phi_1^\dagger\Phi_1
+  \tilde{M}_{H_2}^2\Phi_2^\dagger\Phi_2
+  \right. \nonumber\\
 &\left.
+  \tilde{M}_{H_3}^2\Phi_3^\dagger\Phi_3
  +\tilde{M}_{H_4}^2\Phi_4^\dagger\Phi_4
+  \tilde{M}_{+}^2 \omega_1^+\omega_1^-
+  \tilde{M}_{-}^2 \omega_2^+\omega_2^-
 \right\}
 - \left\{
 (A_u)^{ij} \tilde{u}_{Ri}^\ast \Phi_2 \cdot \tilde{q}_{Lj}
  +\right. \nonumber\\
 &\left.
+(A_d)^{ij} \tilde{d}_{Ri}^\ast \Phi_1 \cdot \tilde{q}_{Lj}
+(A_e)^{ij} \tilde{e}_{Ri}^\ast \Phi_1 \cdot \tilde{\ell}_{Lj}
+(A_1) \omega_1^+ \Phi_1 \cdot \Phi_3 
+(A_2) \omega_2^- \Phi_2 \cdot \Phi_4 + {\rm h.c.} 
 \right\}. 
\end{align}
From $W$ and $\mathcal{L}_{\text{soft}}^{}$, the Lagrangian is constructed as 
\begin{align}
\mathcal{L}=&
 -\left(\frac{1}{2}\frac{\partial^2 W}{\partial \varphi_i\partial\varphi_j}
\psi_{Li}\cdot\psi_{Lj}+h.c.\right)
 -\left|\frac{\partial W}{\partial \varphi_i}\right|^2-\frac{1}{2}(g_a)^2
(\varphi^*_{\alpha}T^a_{\alpha\beta}\varphi_{\beta})^2+\mathcal{L}_{\rm soft}
\;
 ...
 ,
\end{align}
where $\varphi_i$ and $\psi_{L i}$ are respectively
scalar and fermion components of chiral superfields, and    
$T_{\alpha\beta}^a$ and $g_a$ represent generator matrices for
the gauge symmetries and corresponding gauge coupling constants.

The scalar component doublet fields $\Phi_i$ are parameterized as 
\begin{align}
 \Phi_{1,3} = \left[\begin{array}{c}
      \frac{1}{\sqrt{2}}(\varphi_{1,3} + h_{1,3}+ i a_{1,3}) \\
      \phi_{1,3}^-
                    \end{array} \right], 
 \Phi_{2,4} = \left[\begin{array}{c}
      \phi_{2,4}^+ \\
      \frac{1}{\sqrt{2}}(\varphi_{2,4} + h_{2,4}+ i a_{2,4}) 
                    \end{array} \right], 
 \end{align}
where $\varphi_{i}$ are classical expectation values,
$h_i$ are CP-even, $a_i$ are CP-odd and $\phi_i^\pm$ are charged
scalar states.
where $\varphi_{i}$ are classical expectation values,
$h_i$ are CP-even, $a_i$ are CP-odd and $\phi_i^\pm$ are charged
scalar states.
We use the effective potential method to explore the Higgs sector.
At the tree level, the effective potential for the Higgs fields is
given by 
\begin{eqnarray}
V_0
\aneq \sum_{a=1}^4\frac{1}{2}\bar{m}_a^2\varphi_a^2
	+\frac{1}{2}(B\mu \varphi_1\varphi_2+B'\mu' \varphi_3\varphi_4+{\rm
	h.c.})
    +\frac{g^2+g'^2}{32}(\varphi_1^2-\varphi_2^2+\varphi_3^2-\varphi_4^2)^2.
\end{eqnarray}
Using the effective potential, the vacuum is determined 
by the stationary condition as
\begin{align}
\left. \frac{\partial V_{\rm eff}}{\partial \varphi_i}
 \right|_{\langle\varphi_i\rangle=v_i} =0.
 \label{VeffVac}
\end{align}
We assume that the $Z_2$ odd Higgs bosons do not have the vacuum
expectation values (VEVs), 
and we set
$\sqrt{v_1^2+v_2^2} \equiv v$ ($\simeq 246$~GeV) and introduce $\tan\beta=v_2/v_1$. 
At the tree level, $v_3 = v_4 = 0$ is guaranteed by requiring the nonnegative eigenvalues of 
$(\partial^2 V_0/\partial \varphi_i\partial \varphi_j)_{\varphi_{i,j}=0}~(i,j=3,4)$, i.e., 
$\bar{m}_3^2\bar{m}_4^2-B'^2\mu'^2\geq 0,\quad \bar{m}_3^2+\bar{m}_4^2\geq0.$
In the following, we exclusively focus on the $(\varphi_1, \varphi_2)$ space.
For the $Z_2$ even scalar states, 
we have five physical states as in the MSSM; i.e.,
two CP-even $h$ and $H$, a CP-odd $A$ and a pair of charged $H^\pm$
scalar bosons. 
The tree level mass formulae for these scalar states coincide with those
in the MSSM.

We here focus on the one-loop contribution
since radiative corrections on the Higgs sector
are very important to study the EWPT.
The vacuum at the one-loop level is also determined from Eq.~(\ref{VeffVac}) with
the one-loop corrected effective potential.
The one-loop correction to the effective potential at zero temperature is 
given by
\begin{eqnarray}
V_1(\varphi_1,\varphi_2)
\aneq \sum_ic_i\frac{\bar{m}_i^2}{64\pi^2}
 \left(
 	\ln\frac{\bar{m}_i^2}{M^2}-\frac{3}{2}
 \right),
\end{eqnarray}
where $V_1$ is regularized in the $\overline{\rm DR}$-scheme,
$c_i$ is the degrees of freedom of the species $i$, $M$ is a renormalization scale
which will be set on $m^{\rm pole}_t$.
For the zero temperature $T=0$, the one-loop corrected
mass matrix for the CP even neutral bosons
can be calculated from the effective potential.
We here consider the simple case such that $B^{\prime}=B_{\Omega}=\mu^{\prime}=\mu_{\Omega}=0$
in order to switch off the mixing effects.
The renormalized mass of the lightest Higgs boson $h$ is calculated for $m_A \gg m_Z$ as 
\begin{align}
  m_h^2 \simeq m_Z^2 \cos^22\beta + \mbox{(MSSM-loop)} 
 + \frac{\lambda_1^4v^2 c_\beta^4}{16\pi^2}
    \ln \frac{m_{\Omega_2^\pm}^2 m_{\Phi_2^{\prime\pm}}^2}{m_{\bar{\chi}'^\pm_1}^4}  
 + \frac{\lambda_2^4v^2 s_\beta^4}{16\pi^2}
    \ln \frac{m_{\Omega_1^\pm}^2 m_{\Phi_1^{\prime\pm}}^2}{m_{\bar{\chi}'^\pm_2}^4},
 \end{align}
at the leading $\lambda_{1,2}^4$ contributions,
where the one-loop contribution in the MSSM is mainly from the top and stop loop diagram\cite{mh-MSSM}.

Now we quantify the magnitude of the radiative corrections of the $Z_2$-odd particles on $m_h$.
The input parameters are fixed as follows~\cite{kss}:
\begin{eqnarray}
& \tan\beta=3, m_{H^\pm}=500~{\rm GeV}; \nonumber\\
&\tilde{M}_{\tilde{q}}=\tilde{M}_{\tilde{b}}=\tilde{M}_{\tilde{t}}=1000~{\rm
GeV},
 \mu=M_2=2M_1=200~{\rm GeV},
 A_t=A_b=X_t+\mu/\tan\beta; \nonumber\\
& \lambda_1 = 2, \mu'=\mu_\Omega=B_\Omega=B'=0,
 \overline{m}_+^2=\overline{m}_3^2=(500\hspace{2mm} {\rm GeV})^2,
 \overline{m}_-^2=\overline{m}_4^2=(50\hspace{2mm} {\rm GeV})^2.
\label{Param}
\end{eqnarray}
We note that $m_{\Phi_1^{\prime\pm}} < m_{\Phi_2^{\prime\pm}}$ and
$m_{\Omega_1^{\pm}} < m_{\Omega_2^{\pm}}$ in this case.  
On this parameter set, $m_{\Phi_1^{\prime \pm}}^2$, $m_{\Omega_1^{\pm}}^2$ and 
$m_{\tilde{\chi}'^{\pm}_2}^2$ get a significant contribution from $\lambda_2$.
Then their masses become larger for the greater value of $\lambda_2$.
Since the mass parameters $\bar{m}_4^2$ and $\bar{m}_-^2$ are taken to
be small, large 
mass values of $m_{\Phi_1^{\prime \pm}}$ and $m_{\Omega_1^{\pm}}$
 yield the large nondecoupling effects. 
\begin{wrapfigure}{r}{0.5\columnwidth}
\centerline{\includegraphics[width=0.45\columnwidth]{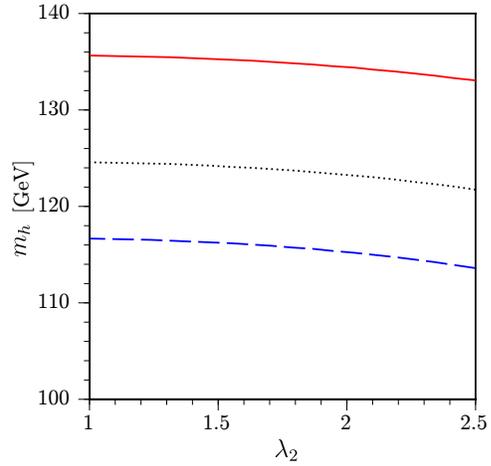}}
\caption{The $Z_2$-even lightest Higgs boson mass as a function of $\lambda_2$.
From the top to the bottom, $X_t/\tilde{M}_{\tilde{q}}=2.0, 1.2$ and 0.6.}\label{fig:mh}
\end{wrapfigure}

\noindent
Fig.~\ref{fig:mh} shows the predicted value of
$m_h$ as a function of $\lambda_2$
varying $X_t/\tilde{M}_{\tilde{q}}=2.0, 1.2$ and $0.6$ from the top to the bottom~\cite{kss}.
We can see that $m_h$ monotonically decreases as $\lambda_2$ increases, which is 
in contrast with the top/stop loop effects.

The coupling constants $\lambda_1$ and $\lambda_2$ are
free parameters of the model. Its magnitude, however,
is bounded from above by the condition that there is no
Landau pole below the given cutoff scale $\Lambda$.
As we are interested in the model where the first order
EWPT is sufficiently strong, we allow rather
larger values for these coupling constants, and do not
require that the model holds until the grand unification scale. 
A simple renormalization group equation analysis tells us that
for assuming $\Lambda = 2$~TeV, $10$~TeV or $10^{2}$~TeV,
the coupling constant can be taken to be at most $\lambda_2 \sim 2.5$, $2.0$
or $1.5$, respectively.

\section{Electroweak Phase Transition}
 
The nonzero temperature effective potential is 
\begin{align}
V_1(\varphi_1,\varphi_2;T) = \sum_ic_i\frac{T^4}{2\pi^2}I_{B,F}\left(\frac{\bar{m}_i^2}{T^2}\right),
 \end{align}
where $B(F)$ refer to boson (fermion) and $I_{B,F}$ take the form
\begin{eqnarray}
I_{B,F}(a^2) = \int_0^\infty dx~x^2\ln\Big(1\mp e^{-\sqrt{x^2+a^2}}\Big).\label{IBF}
\end{eqnarray}
Since the minimum search using $I_{B,F}$ is rather time-consuming, 
we will alternatively use the fitting functions of them that are employed in Ref.~\cite{Funakubo:2009eg}. 
 
For an electroweak baryogenesis scenario
to be successful, the sphaleron rate in the broken phase 
should be smaller than the Hubble constant. This condition is translated into 
\begin{eqnarray} 
\frac{v_C}{T_C}=\frac{\sqrt{v_1^2(T_C)+v_2^2(T_C)}}{T_C}\gsim\zeta,
\label{sph_dec}
\end{eqnarray}
where $T_C$ is the critical temperature, $v_C$ is the Higgs VEV at $T_C$, and
$\zeta$ is a $\mathcal{O}(1)$ parameter.
To obtain $\zeta$ within a better accuracy, 
the sphaleron energy and zero-mode factors of the fluctuations around the sphaleron 
must be evaluated. 
In the SM, the sphaleron energy is simply a function of the Higgs boson mass.
As the Higgs boson becomes heavier, the sphaleron energy gets larger as well~\cite{Klinkhamer:1984di},
leading to the smaller $\zeta$~\cite{Arnold:1987mh}. 
In this model, on the other hand, $\zeta$ depends on more parameters.
For simplicity, we here take $\zeta=1$, which is often adopted as a rough criterion in the literature.

In our analysis, $T_C$ is defined as the temperature at which the effective potential has the
two degenerate minima. We search for $T_C$ by minimizing 
\begin{eqnarray}
V_{\rm eff}(\varphi_1, \varphi_2; T) = V_0(\varphi_1, \varphi_2)
+V_1(\varphi_1, \varphi_2)+V_1(\varphi_1, \varphi_2; T), 
\end{eqnarray}
where the field-dependent masses are modified by adding thermal
corrections.

\begin{figure}[t] 
\begin{center}
   \epsfig{file=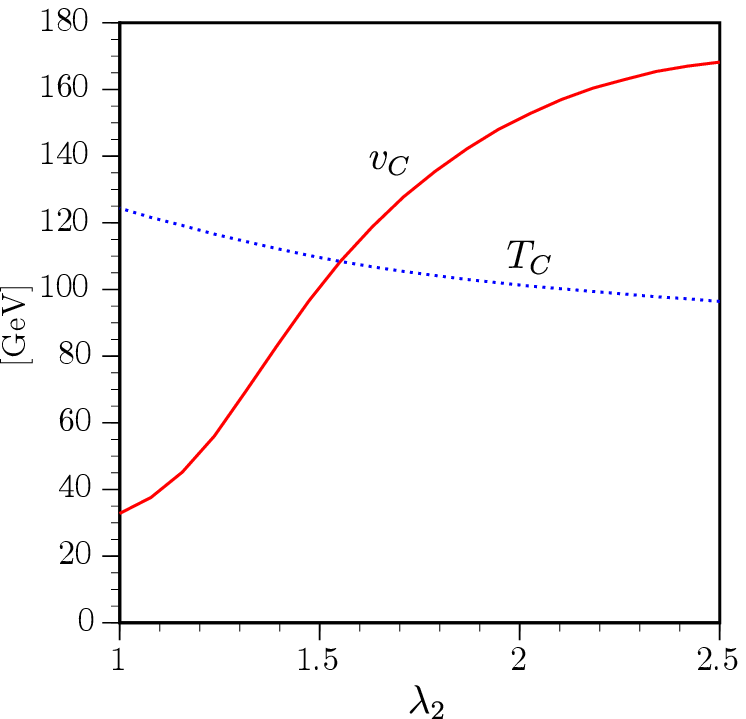,width=6.cm}
   \epsfig{file=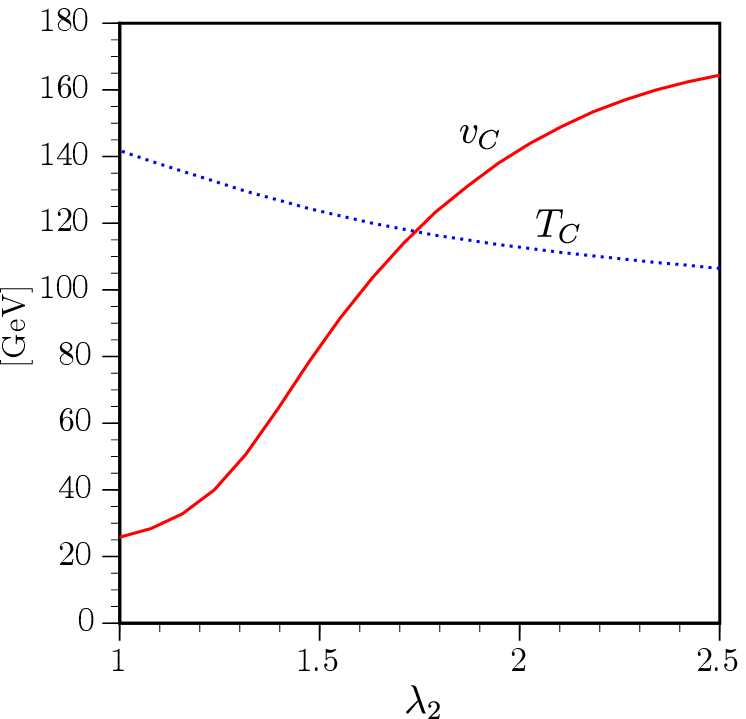,width=6.cm}
\caption{(Left) $v_C$ and $T_C$ vs. $\lambda_2$ with $X_t/\tilde{M}_{\tilde{q}} = 0.6$. 
The other input parameters are the same as in the 
Fig.~\ref{fig:mh}. The sphaleron decoupling condition (\ref{sph_dec})
 can be satisfied for $\lambda_2\gtsim 1.6$.
(Right) $v_C$ and $T_C$ vs. $\lambda_2$ with $X_t/\tilde{M}_{\tilde{q}} = 2.0$. 
The other input parameters are the same as in the 
Fig.~\ref{fig:mh}.
 The sphaleron decoupling condition (\ref{sph_dec})
 can be satisfied for $\lambda_2\gtsim 1.8$.}
 \label{fig:PT_light}
\end{center}
\end{figure}

In Fig.~\ref{fig:PT_light} (Left), $v_C$ and $T_C$ are plotted as a function of $\lambda_2$
in the light $h$ scenario $(X_t/\tilde{M}_{\tilde{q}}=0.6)$~\cite{kss}:
see Fig.~\ref{fig:mh}.  
The sphaleron decoupling condition (\ref{sph_dec}) can be fulfilled for $\lambda_2\gtsim 1.6$ 
due to the nondecoupling effects coming from $\phi'^\pm_1$ and $\Omega_1^\pm$.

We also evaluate $v_C$ and $T_C$ in the heavy $h$ scenario
$(X_t/\tilde{M}_{\tilde{q}} = 2.0)$ as shown in Fig.~\ref{fig:PT_light} (Right)~\cite{kss}.
The sphaleron decoupling condition can be satisfied for $\lambda_2\gtsim1.8$.
Though the parameter region is a bit narrower than the light Higgs scenario, 
the lightest Higgs boson mass as large as 134~GeV is still consistent with 
the decoupling condition.

\section{Phenomenological predictions and Discussions}

The nondecoupling effect of the extra $Z_2$ odd charged scalar bosons
on the finite temperature effective potential is an essentially important
feature of our scenario in order to realize strong first order phase transition.
The same physics affects the triple Higgs boson coupling with
a large deviation from the SM (MSSM) prediction 
as discussed in Ref.~\cite{ewbg-thdm2}. 
Such deviation in the triple Higgs boson coupling can be
15-70 \% \cite{KOSY,ksy}, and we expect
that they can be measured at the future linear collider such as
the ILC or the CLIC.

In our model, in order to realize the nondecoupling
effect large, the invariant parameters $\mu^\prime$ and $\mu_{\Omega}$
are taken to be small. Consequently, the masses of extra charginos
are relatively as light as 100-300~GeV. 

 The several comments on the current analysis are in order.
I) In the MSSM, it is found that $\zeta\simeq 1.4$~\cite{Funakubo:2009eg},
which is 40\% stronger than one we impose in our analysis.
We emphasize that even if we take $\zeta=1.4$ in our model 
a feasible region still exists for the relatively large $\lambda_2$, for example, 
$\lambda_2\gtsim2.2$ even in the heavy $h$ case.
The cutoff scale $\Lambda$ is  still around the multi-TeV scale. 
II)     
As in the MSSM, the strength of the first order EWPT can get enhanced
if the (almost) right-handed stop is lighter than the top quark, enlarging the possible region.
III) As in the MSSM, the charginos and the neutralinos can play an essential role in
generating the CP violating sources as needed for the bias of the
chiral charge densities around the Higgs bubble walls.
  The $Z_2$ odd charginos $\widetilde{\chi}'^\pm_{1,2}$ may
     also be helpful.
IV) We have confirmed that
the charge and the $Z_2$ are not broken at the tree level for our
     parameter set.
V) The potential analysis beyond the tree level 
will be a future problem.
VI)
 If the $Z_2$ symmetry is exact, the lightest (neutral) $Z_2$ odd particle
 can be a candidate of dark matter.
     If it is the scalar boson, its properties would be similar to
     those of the inert doublet model~\cite{Barbieri:2006dq}.
 A $Z_2$ odd neutralino may also be a candidate
 for dark matter.  

\section{Conclusions}

 We have found that the nondecoupling loop effects of additional charged
 scalar bosons can make the first order EWPT strong enough
 to realize successful electroweak baryogenesis. 
 We conclude that this model is a good candidate
 for successful electroweak baryogenesis. 

 \section*{Acknowledgments}

The speaker was supported by a JSPS Grant in Aid for Specially Promoted
Program ``A Global Research and Development Program of a
State-of-the-Art Detector System for ILC''.


\begin{footnotesize}


\end{footnotesize}


\end{document}